\pdfoutput=1
\documentclass[preprint, review, 11.5pt]{elsarticle}
\usepackage{lineno,hyperref}
\usepackage{todonotes}
\usepackage{multirow}
\usepackage{longtable}
\usepackage{appendix}
\usepackage{graphicx}
\usepackage[]{xcolor}
\usepackage{url}

\usepackage{etoolbox}
\makeatletter
\AtBeginDocument{\def\paragraph{\def\@period{\upshape:}%
    \vskip \@startsection{paragraph}{4}{\z@}{6\p@ \@plus \p@}% % GM and Wolfgang May - 11/30/06
    {-5\p@}{\subsecfnt}%
}}
\journal{Software Impacts}
\begin{document}
\begin{frontmatter}

\title{OBAMA, an Ontology-Based Software Tool for Agile Method Adoption}

\author[1]{Soreangsey Kiv}
\author[3]{Yves Wautelet}
\author[2]{Samedi Heng}
\author[1]{\\Manuel Kolp}

\address[1]{LouRIM, UCLouvain, Louvain-La-Neuve, Belgium \\ \{soreangsey.kiv, manuel.kolp\}@uclouvain.be}
\address[2]{HEC Li\`ege, Universit\'e de Li\`ege, Li\`ege, Belgium\\ 
samedi.heng@uliege.be}
\address[3]{KULeuven, Leuven, Belgium\\
\{yves.wautelet@kuleuven.be}

\begin{abstract}
Tools like Prot\'eg\'e support the creation and edition of one or more ontologies in a single workspace. They nevertheless require a user to be familiar with this kind of abstractions and their supporting techniques such as a reasoner and SPARQL queries. This paper presents a step-by-step implementation of a prototype-tool that allows retrieving and displaying easily the information about agile practices contained in an ontology using Python programming language. Future development includes the flexible insertion, modification, and removal of knowledge by the user.
\end{abstract}

% %%Research highlights
\end{frontmatter}

%%%%%%%%%%%%%%%%%%%% author.tex %%%%%%%%%%%%%%%%%%%%%%%%%%%%%%%%%%%
%
% sample root file for your "contribution" to a proceedings volume
%
% Use this file as a template for your own input.
%

\newpage
\noindent
\textbf{Code metadata}\\

%\begin{table}[!h]
\noindent
\begin{tabular}{p{5.5cm}p{6.5cm}}
\hline

Current code version & \textit{v1.0} \\
\hline
Permanent link to code/repository used for this code version &  https://github.com/soreangsey/ontology.git\\
\hline
Permanent link to Reproducible Capsule &  \textit{ doi.org/10.24433/CO.8008250.v1} \\
\hline
Legal Code License   & \textit{MIT license} \\
\hline
Code versioning system used & \textit{None} \\
\hline
Software code languages, tools, and services used & \textit{Python, owlready2, wxPython, Web Ontology Language (OWL)}\\
\hline
Compilation requirements, operating environments \& dependencies & \textit{ doi.org/10.24433/CO.8008250.v1} (See requirements.txt)
\\
\hline
If available Link to developer documentation/manual & \textit{None}\\
\hline
Support email for questions & \textit{$soreangsey.kiv@uclouvain.be$}  \\
\hline
\end{tabular}\\

\section{Introduction}
\label{introduction}
A lot of different practices are grouped under the umbrella of agile software development. Being agile does not necessarily mean adopting all of these practices \citep{eilers2020doing,rahman2018doing}. Organizations or individual software development teams often select a method (like Scrum, XP, SAFe, etc.) and try to adapt it in a specific business context. Then, all the practices prescribed by the chosen method do then not need to be implemented in the specific business environment and a customized on-demand implementation is generally preferable \citep{abbas2010using,kiv2021towards}.

The Ontology-Based for Agile Method Adoption (OBAMA) tool\footnote{A demo in the form of a video of the tool can be found at \url{https://bit.ly/3B5dcJN}.} is a software system that aggregates a lot of knowledge on the adoption of agile practices found within the scientific literature. First, it fully implements an ontology depicting the relevant concepts in the field of systematic (agile) practices adoption built out of a systematic literature review. Second, it hosts a lot of knowledge gathered from literature reporting on the experience in adopting specific agile practices. Finally, through SPARQL queries, practitioners can easily access the most relevant information when adopting specific agile practices. The process to retrieve relevant information from the ontology using available tools such as Protégé tool requires users to follow some steps including (1) creating a new instance to represent a development team, (2) describing the team’s properties (3) executing the reasoner to automatically link the team instance to the other instances based on inference rules, (4) and using SPARQL to retrieve the desired concerns. Since this information retrieval process is complicated and requires a user to have some preliminary knowledge, a user-friendly tool to simplify it is deemed important to build.

%the problem with protege tool and the purpose of the paper

The tool presented in this paper is part of a larger research effort to build and validate an ontology for agile practice adoption. A primary ontology for practices adoption was presented in \cite{DBLP:conf/icsoft/KivHKW17,DBLP:conf/icsoft/KivHWK17,DBLP:conf/xpu/KivHKW19} while the entire research to build the ontology validating and populating it has been depicted in \citep{kiv2022using}. Ontology creation has been used previously in the domain of agile methods like in \cite{DBLP:conf/caise/WauteletHKM14} with the aim of creating requirements (conceptual) models and assist in software engineering (see \cite{DBLP:journals/cl/WauteletHKK17}); it is here used as a database structure to support the knowledge retrieval of agile practice adoption in a flexible and custom manner. This paper serves as a reference for the OBAMA tool, it describes the technology supporting the software as well as its architecture and impact.

%structure of the paper
In Section \ref{important components}, we provide a basic explanation about each of the important components needed to develop the tool. Section \ref{tool architecture} explains the technical architecture of our tool, with a set of steps from loading until displaying the information to the users. Section \ref{tool functionality} describes the functionality of the tool. Finally, we conclude with the result, the impact and future research directions in Section \ref{conclusion}
\section{Software Description}
\subsection{Main Technologies Supporting the OBAMA-Tool}
\label{important components}

There are four main technologies used for the development of our supporting tool are: 
\begin{itemize}
\item An ontology file written in \emph{Web Ontology Language} (\emph{OWL}) \citep{mcguinness2004owl}. This file constitutes the database for storing the collected knowledge about agile practice adoption;
\item The Python programming language was chosen because it provides an extensive \emph{Application Programming Interface} (\emph{API}) to easily work with the ontology;
\item \textit{Owlready2}, an ontology-oriented programming package in Python that allows loading, modifying, saving, and executing a reasoner on the ontology \citep{lamy2017owlready}. Since the HermiT OWL reasoner included in \textit{Owlready2} is written in Java programming language, we need to install Java.exe to execute it;
\item \textit{wxPython}, a cross-platform GUI toolkit that we use to create the interface \citep{precord2015wxpython}. 
\end{itemize}

These technologies are described in this section.

\subsubsection{The Ontology File Written in Web Ontology Language}

All the descriptions of concepts, relationships, and inserted knowledge related to agile practice adoption are written in the OWL format. The ontology file can be seen as the database file of the system.

OWL is an ontology language recommended by the World Wide Web Consortium (W3C) and it provides three sub-languages: \emph{OWL Lite}, \emph{OWL Description Logic} (\emph{DL}), \emph{OWL Full} \citep{mcguinness2004owl}. OWL Lite is the core language for creating an ontology with a minimum number of restrictions and constraints. OWL DL is a restricted version of OWL Full. It has some restrictions on the context in which language construction can be used to ensure decidability. OWL Full uses the same vocabulary as the above presented OWL DL version. However, it is much more expressive and has fewer restrictions than other versions. We use OWL Full for our ontology creation as it is more descriptive and less restrictive, which makes it the most convenient choice for usage. OWL can be seen as an extension of the \emph{Resource Description Framework} (\emph{RDF}) vocabulary because each OWL document is an RDF. Like RDF, OWL vocabulary includes a set of XML syntaxes that are used to describe knowledge in triples, ``Subject-Predicate-Object'', just like a statement in a natural language. For example, within the RDF format statement \emph{``practice is composed of an activity''}: \emph{practice} is the subject, \emph{is composed of} is the predicate, and \emph{activity} is the object.

\begin{figure*}

\centering

 \includegraphics[width=1\textwidth]{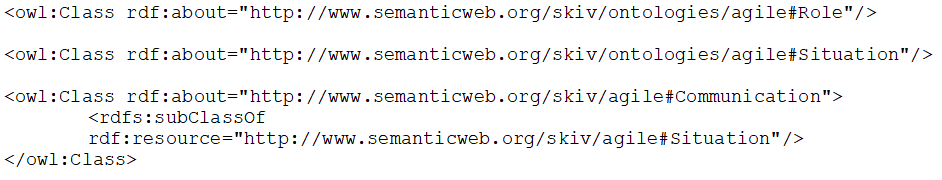}

 \caption{Example of OWL syntax to create classes.}
\label{fig:syntaxClass}
\end{figure*}

Following the procedure of ontology creation \citep{noy2001ontology}, we started by defining the key concepts (classes) and then hierarchically refined the concepts into sub-classes. In the concept refinement process, one of the decisions to take was whether we should introduce a new sub-class or simply distinguish through instances with different property values \citep{noy2001ontology}. For instance, there are seventeen different types of \textit{situation} that can affect the agile adoption result, including \textit{team size}, \textit{project size}, \textit{type of communication}, \textit{team distribution}, etc. In this case, we need to decide whether we should represent each type of \emph{situation} by a sub-classe or an instance. According to \citep{noy2001ontology}, we can identify a class when each type has a different relationship with other class(es). Based on this principle, as each type of \textit{situation} has a different effect (\textit{hurt} or \textit{harm}) on the \textit{requisite}, we represent each type of \textit{situation} by a sub-class in our ontology model. An example of OWL syntax to create class and sub-class in our ontology is illustrated in Figure \ref{fig:syntaxClass}. In the example, \textit{Role} and \textit{Situation} are the main classes where \textit{Communication} is the sub-class of \textit{Situation}.

\begin{figure*}
 
\centering

 \includegraphics[width=1\textwidth]{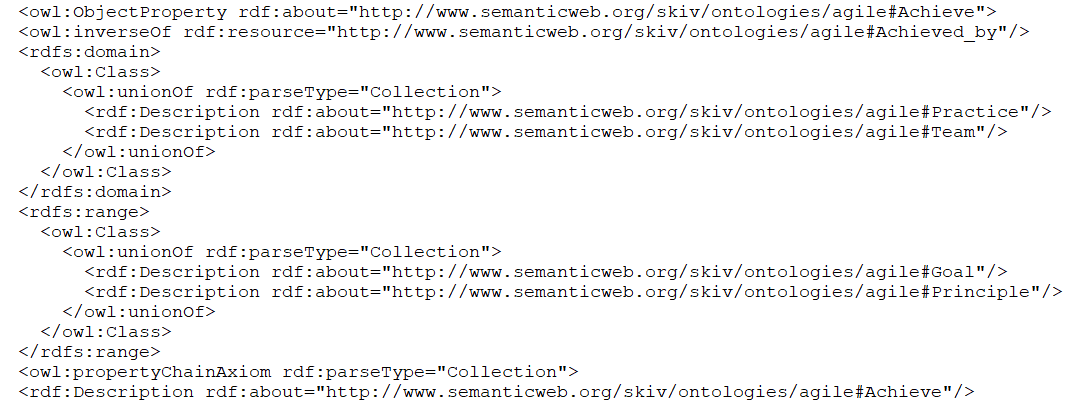}

 \caption{Example of OWL syntax to define object property.}
\label{fig:syntaxObject}
\end{figure*}

Once all classes were created, we continued with the definition of their properties that include two types: data and object. (1) The data property is used to link the individuals (i.e., instances) and data value. Every class in our model has only two data properties (i.e., Name and Description). Based on the literature, these two data properties are the only common elements used to describe each class by agile practitioners. These properties are String types. (2) Object property is used to link individuals (i.e., instances) and individuals (i.e., instances). Both links are built in the form of ``Domain - data/object property - Range''. For instance, the relationship ``Practice - Achieve - Goal'' has \textit{Achieve} as an object property where \textit{Practice} is its domain and \textit{Goal} is its the range. Figure \ref{fig:syntaxObject} is an example of OWL syntax to define an object property. In the example, the object property \textit{Achieve} is the reverse of another object property named \textit{Achieved by}. In addition, either a \textit{team} or \textit{practice} (domain) can \textit{achieve} either \textit{goal} or \textit{principle} (range). 

\begin{figure*}
 
\centering

 \includegraphics[width=1\textwidth]{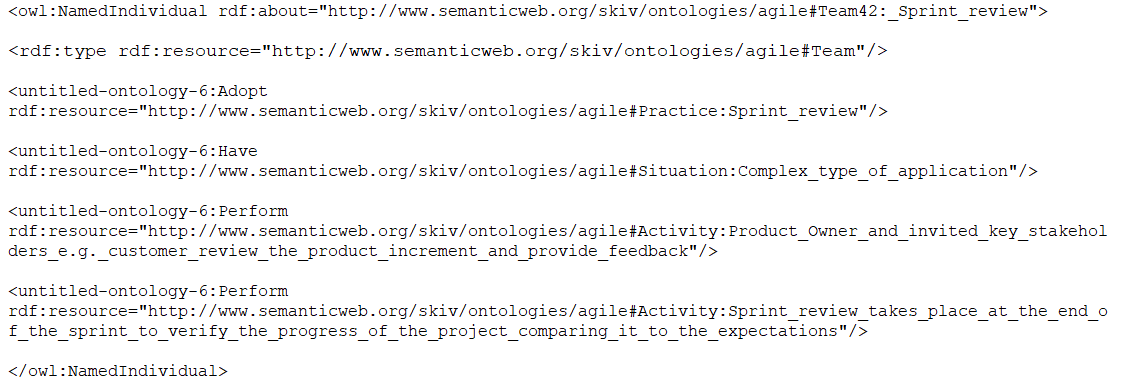}

 \caption{Example of OWL syntax to create an individual and define its relationships.}
    \label{fig:syntaxIndividual}
\end{figure*}

Finally, we created the individuals (i.e., instances) based on the knowledge we had extracted from the literature. We started by creating an individual of a class \textbf{Team} for each case study. We then continued to create other individuals for the other information related to that team, for instance, the \textit{practice} the team adopted and the \textit{activity} it performed as part of the practice, etc. When all the individuals were created, we built their relationships by defining their object properties. Figure \ref{fig:syntaxIndividual} is an example of OWL syntax to create an individual \textit{Team42: Sprint review} and define its relationships with other individuals.

\subsubsection{SPARQL}

\begin{figure*}

\centering

 \includegraphics[width=0.9\textwidth]{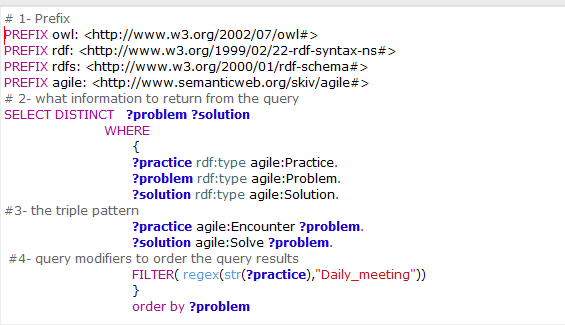}

 \caption{Example of SPARQL query.}

 \label{fig:sparql_query}

\end{figure*}

SPARQL is a query language that was standardized in 2008 by the W3C that is used for querying RDF data \citep{prud2005sparql}. Most forms of SPARQL queries contain a set of basic triple graph patterns (subject, predicate, object) where each pattern matches a sub-graph of RDF data and the terms from that sub-graph may be substituted for the variables \citep{schmidt2010foundations}. To query for the information in our ontology, each SPARQL query comprises 4 parts of information in the following order. (1) A prefix that is the abbreviation of the \emph{Uniform Resource Identifiers} (\emph{URIs}), (2) the information to return from the query, (3) the triple patterns, and (4) the query modifiers to filter and order the query results. Figure \ref{fig:sparql_query} is an example of how to write a SPARQL query to select all the \textit{solutions} to \textit{solve} the \textit{problems} that are \textit{encountered} by \textit{Daily meetings}.

\subsubsection{OWLReady2}

\begin{figure*}

\centering

 \includegraphics[width=0.9\textwidth]{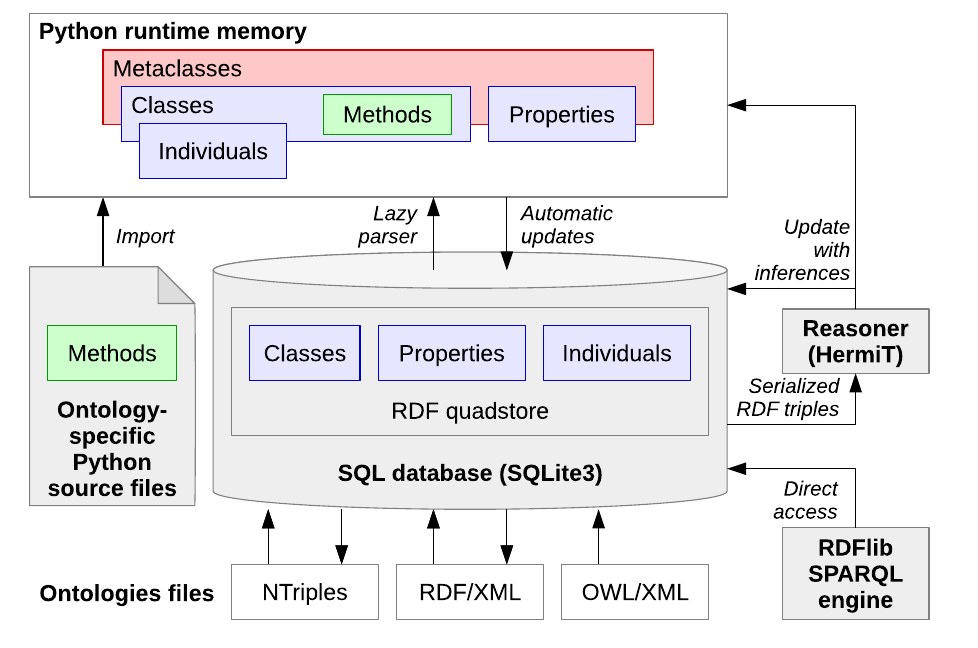}

 \caption{OWLReady2 Architecture \citep{lamy2017owlready}}

 \label{fig:owlready2Archi}

\end{figure*}

Owlready2 is a module for ontology-oriented programming in Python that we use to access the ontology, execute the reasoner, and query the information. The general structure of the Owlready2 is made of five main components as shown in Figure \ref{fig:owlready2Archi}. They include (1) an optimized RDF quadstore implemented with an SQL database (i.e., SQLite3) and stored either in memory or on disk in a file, (2) meta-classes for OWL classes and constructs, (3) optional ontology-specific Python source files defining methods to insert into OWL classes, (4) the HermiT OWL reasoner for performing automatic classification and (5) the SPARQL engine from the RDFlib Python module that is used for RDF query \citep{lamy2017owlready}. 

\begin{figure*}

\centering

 \includegraphics[width=0.99\textwidth]{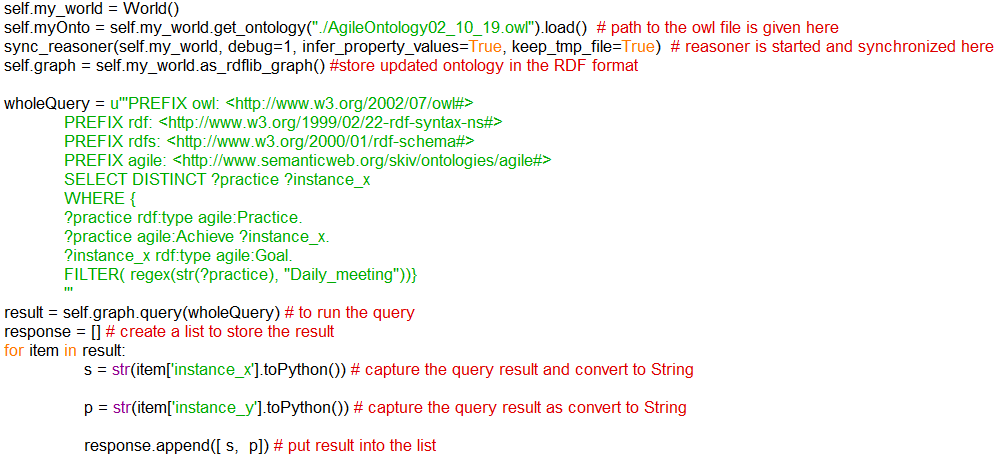}

 \caption{Ontology query in Python programming language.}

 \label{fig:pythonQuery}

\end{figure*}

The general algorithm that we use to query the ontology is OWLReady2 as shown in Figure \ref{fig:pythonQuery}. First, the ontology is imported into \textit{RDF quadstore} and their concepts and/or entities are saved in Python objects. Then, the HermiT reasoner is executed. After that, the updated entities of the ontology are automatically stored in a memory in RDF format. We then perform SPARQL queries using the RDFlib graph. Finally, a Python function captures and converts the results into Strings and stores them in a list-type variable.

\subsubsection{wxPython}

wxPython is a cross-platform toolkit that is used to create desktop \emph{Graphical User Interface} (\emph{GUI}) applications. wxPython provides a set of methods and objects that programmers can use to simply and easily create programs with a robust and highly functional GUI \citep{precord2015wxpython}. Technically, wxPython is a wrapper over a C++ GUI API called wxWidgets which are essential building blocks of a GUI application. wxPython provides a lot of widgets that can be divided into six groups. (1) Base Widgets provide basic functionality for derived widgets, (2) top-level widgets provide different functions independently of each other, (3) containers that contain other widgets, (4) dynamic widgets that can be edited by users, (5) static widgets that are used to only display information but cannot be edited by the users, and (6) random widgets such as status-bar, toolbar, and menu-bar, etc. There is a specific relation among widgets in wxPython that is developed by inheritance. Widgets form a hierarchy; they can inherit functionality from other widgets. The advantages of wxPython include being open-source, cross-platform, easy to write, and easy to understand.

\subsection{Tool Architecture}

\label{tool architecture}

\begin{figure*}

\centering

 \includegraphics[width=1\textwidth]{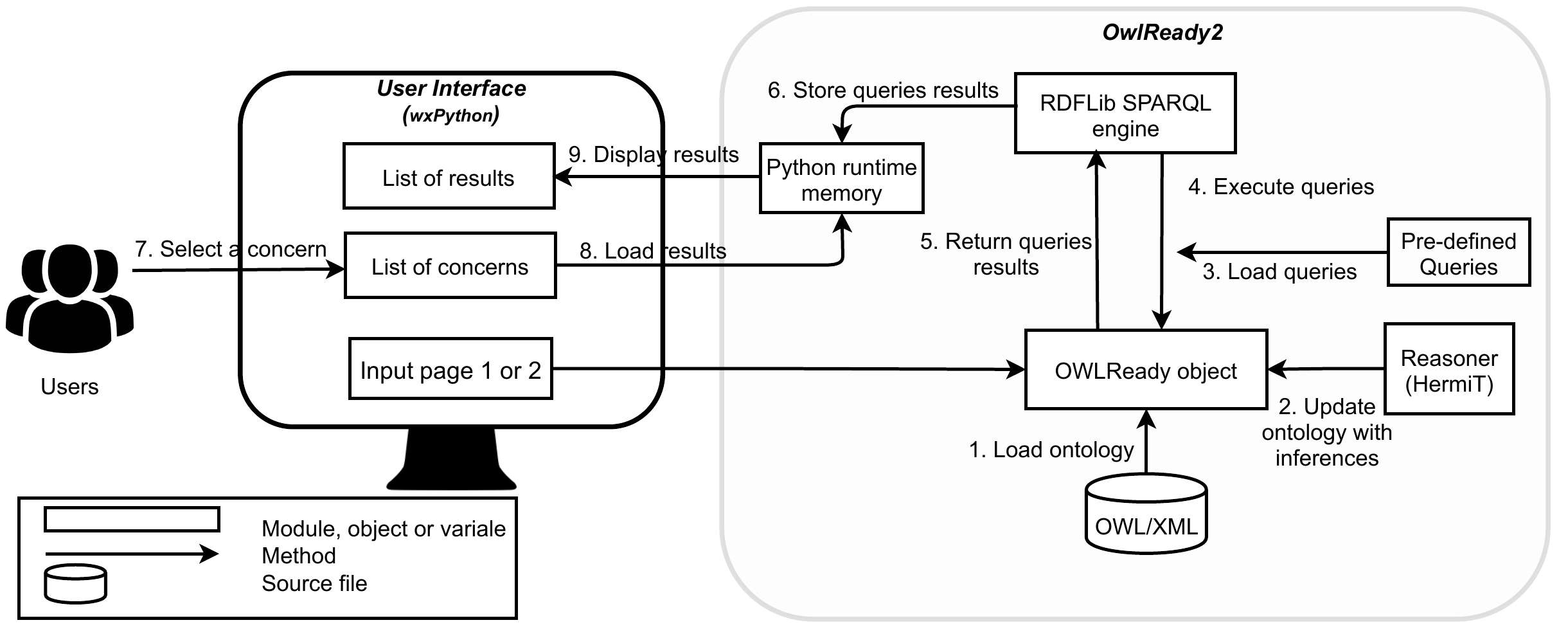}

 \caption{Technical architecture for all the information related to practice.}

 \label{fig:toolArchitecture1}

\end{figure*}

\begin{figure*}

\centering

 \includegraphics[width=1\textwidth]{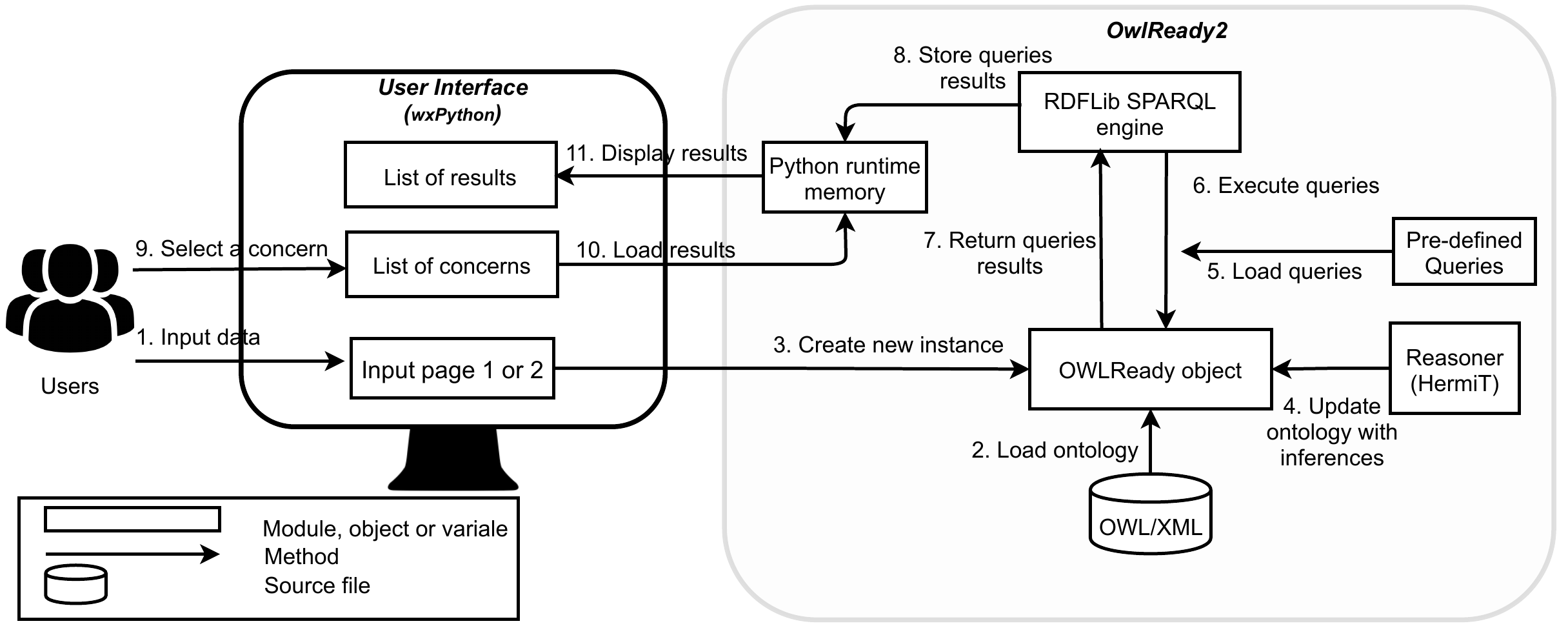}

 \caption{Technical architecture for questions and answers based on the inputs.}

 \label{fig:toolArchitecture2}

\end{figure*}

The architecture of the tool is illustrated in Figures \ref{fig:toolArchitecture1} and \ref{fig:toolArchitecture2} with two different scenarios. One is to retrieve all the information while the other is to retrieve the information based on the user's inputs.

\subsubsection{Technical architecture for all the information related to a practice}

In the scenario where all the information in ontology is retrieved and listed, the retrieval process occurs only the first time the program is executed, and the returned information is then stored in the permanent memory for future usage. As there is no any change to the information, loading the information from the memory saves much more time than re-loading from the ontology file. The program is executed in 9 steps, from loading data from the ontology to displaying the results for the users. 

\begin{enumerate}

    \item The program imports the ontology from the source file (OWL), loads their entities recursively, and wraps them in a Python object named \textit{OwlReady Object};

    \item The HermiT reasoner is then executed to auto-generate the classification of the classes, properties, and individuals based on the inference rules. After that, it updates the \textit{Owlready Object};

    \item The program loads all the pre-written queries to retrieve the information related to all the concerns (one query for one concern) and stores them in a list-type variable;

    \item RDFLib SPARQL engine then iteratively runs all the queries from the list;

    \item The results of each query retrieved by the RDFLib SPARQL engine are then converted and stored in another list-type variable;

    \item Once all the queries are run, all the query results are stored in permanent memory for future usages; 

    \item From the GUI, users select a concern from the list;

    \item The program loads the information related to the selected concern from the memory;

    \item Finally, the information is displayed in table format. 

\end{enumerate}

\subsubsection{Technical architecture for the information related to practice based on inputs}

In the scenario where the information is retrieved based on the users' inputs, the retrieval process occurs each time the users click on the button ``Calculate result'' in the ``Input page 1'' or the ``Input page 2''. There are 11 steps in total from loading data from the ontology to displaying results for the user. 

\begin{enumerate}

    \item From the GUI, users provide the input values (goals or/and situations);
     \item The program imports the ontology source file, loads their entities recursively, and wraps them in a Python object named \textit{OwlReady Object}. This step is done in the previous scenario;

    \item The program then creates a \textit{temporary ontology object} of the class \textit{Team} where the input values are its properties;

    \item The \emph{HermiT} reasoner is then executed to auto-generate the classification of the classes, properties, and individuals based on the inference rules. By doing so, the \textit{temporary ontology object} are now linked to the other individuals based on pre-defined inference rules. After that, it updates the \textit{Owlready Object}; 

    \item The program loads all the  pre-written queries to retrieve the information related to all the concerns (one query for one concern) and stores them in a list-type variable;

    \item The \emph{RDFLib SPARQL} engine then iteratively runs all the queries from the list;

    \item The results of each query retrieved by RDFLib SPARQL engine are then converted into Strings and stored in another list-type variable;

    \item All the query results are stored in a temporary memory;

    \item From the GUI, users select a concern from the list;

    \item The program loads the information related to the selected concern from the memory;

    \item Finally, the information is displayed in the table format. 

\end{enumerate}

\subsection{Functionality of the Tool}
\label{tool functionality}

\begin{figure}[htbp]
\centering
 \includegraphics[width=1\textwidth]{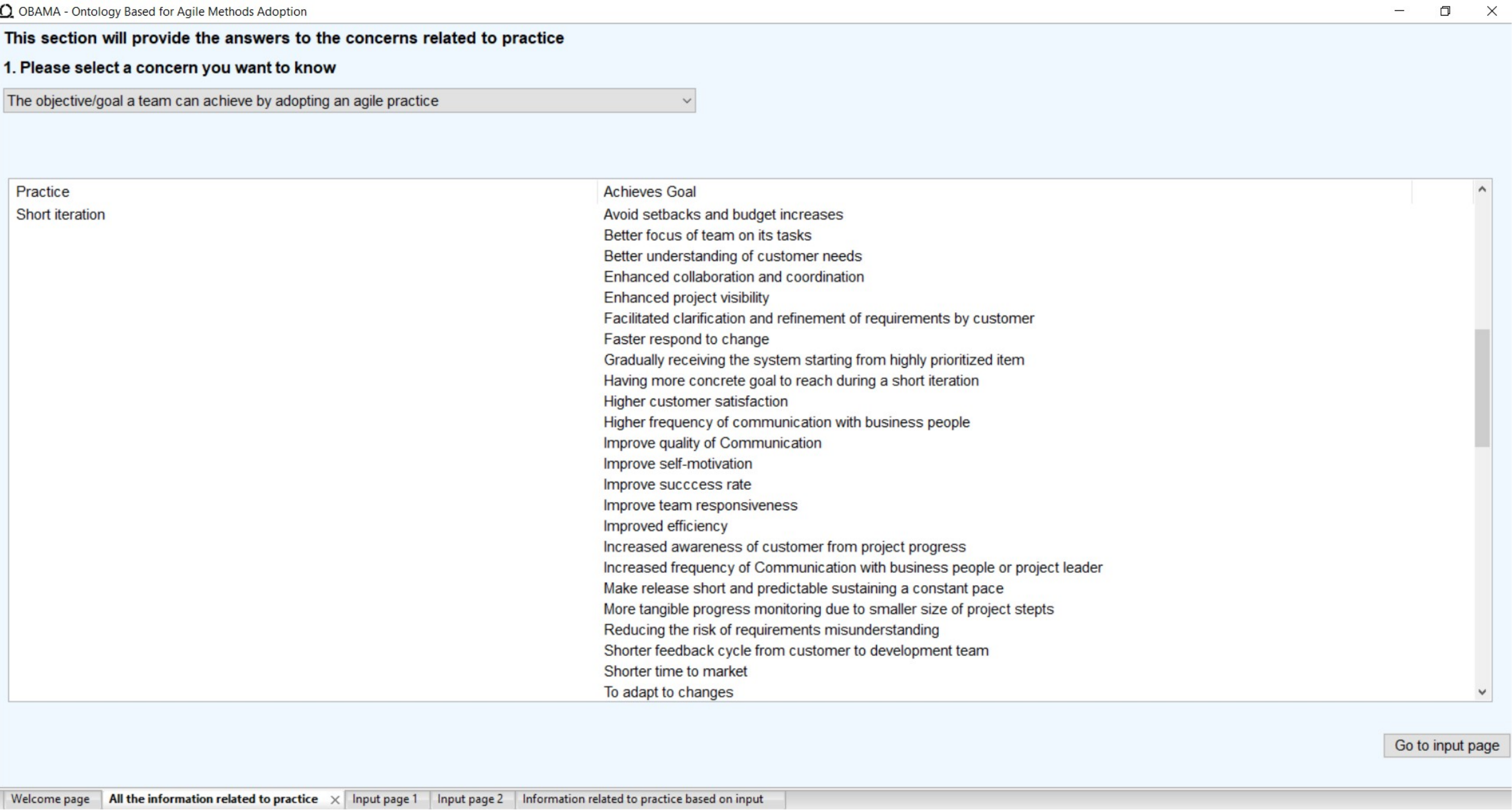}
 \caption{Screenshot of OBAMA tool.}
 \label{fig:toolprintscreen}
\end{figure}

In this paper, we thus focus on the prototype-tool that allows retrieving and displaying information. This tool is meant to help practitioners see how useful the ontology could be when the knowledge is available. The part where a user can insert, modify, and delete knowledge will be developed at a later stage. It was created in a notebook-style where each functionality is put on a different page as shown in Figure \ref{fig:toolprintscreen}. 
\begin{itemize}
    \item The first page, called ``Welcome page", serves to introduce to users the purpose of the tool and some usage guidelines;
    \item The second page ``All the information related to practice" allows users to access all the information related to agile practice adoption that we have inserted. All concerns are listed in a combo-box. By choosing a concern from the list, the information will be displayed underneath in a table format;
    \item If users want to filter the information based on goals, they need to provide input values in the ``Input page 1''. Since most goals in adopting agile methods can be mapped to the Agile Manifesto \citep{kiv2018agile}, we provide on this page a list of agile values and principles from which users can choose;
    \item If users want to filter the information based on their situations, they need to provide their input values in the ``Input page 2''. On this page, we provide 13 situational factors which have been collected through an SLR \citep{campanelli2015agile}. Each of the factors has a list of different values from which users can choose. In both ``Input page 1'' or ``Input page 2'', there is a button ``Calculate result'' that allows starting the data retrieval process;    
     \item Our tool retrieves the information based on users' inputs before it displays the results in the page ``Information related to practice based on input''. Users can find the desired information by choosing a concern from the list.
\end{itemize}

Our ontology file and source code in Python programming language can be found at \citep{obamacode}.

\section{Impact and Conclusion}
\label{conclusion}

Ontologies have been widely used in many fields, but not many research articles address the difficulties of working with ontologies for an average user and how to overcome this problem. The purpose of this paper is to elaborate on how to create a user-friendly tool in Python to help users retrieve information without any required preliminary knowledge. As said, the software is aimed to be used by practitioners, especially novice ones since their potentially limited experience makes the software tool especially relevant for them. By using our tool, practitioners can easily get general information on agile practice adoption. They can also filter for only relevant information based on their goals and context. The tool has been evaluated by agile practitioners on a set of criteria related to efficiency in \cite{kiv2022using}. Results allow us to declare that all the features in our tool are efficient enough for users. Most experts ``Somewhat agree'' or ``Agree'' that they can access, understand and input the information easily. Even though we cannot say it is already satisfying enough for massive adoption by users yet, but this result is overall still reassuring. It is also important to notice that, most experts ``Agree'' that our tool helps them decide whether or not an agile practice is suitable for their team. 

The tool has been developed in a laboratory setting and validated by practitioners on the mentioned elements. More work is currently made with novice agilists in real-life settings to further evaluate the tool but also improve it to align at best to their requirements.

\bibliographystyle{model5_names}
\biboptions{authoryear}
\bibliography{mybibliography}

\newpage

\end{document}